\documentclass{PoS}

\usepackage{latexsym}
\usepackage{amsmath}
\usepackage{epsfig}
\usepackage{graphics}
\PoS{PoS(LAT2005)258}

\title{Is there a third-order phase transition in quenched QCD?}

\ShortTitle{Third-order phase transition in QCD?}

\author{L. Li and \speaker{Y. Meurice}\\

        University of Iowa\\

        E-mail: \email{yannick-meurice@uiowa.edu}}

\abstract{
We discuss the connection  between the contributions of large field configurations 
and the large order behavior of perturbation theory. For quenched $QCD$, the 
sensitivity of the average plaquette to a removal of large field configurations has a narrow peak near $\beta=5.6$.
Various analysis of the order 10 weak coupling series for the
plaquette give robust indications
for a singularity in the third derivative of the free energy (second
derivative of the plaquette) with respect to $\beta$, near $\beta$ = 5.7.
We report results of numerical calculations. 
The peak in the third derivative of the free energy present 
on $4^4$ lattices disappears if the size of the lattice is increased isotropically 
up to a $10^4$ lattice. 
On the other hand, on $4\times L^3$ lattices, it
persists when $L$ increases. The location of the peak coincides with the onset of 
a non-zero average for the Polyakov loop and seems related to the finite 
temperature transition. We also discuss the discrepancy between the perturbative series and 
the numerical values of the plaquette. 
}

\FullConference{XXIIIrd International Symposium on Lattice Field Theory\\

                 25-30 July 2005\\

                 Trinity College, Dublin, Ireland}

\begin{document}
\section{Introduction}
Perturbation theory can be a frustrating tool for field theorists. Sometimes, it provides 
extremely accurate answers, sometimes it is not even qualitatively correct. In recent 
years, our main goal has been to construct 
modified perturbative series which are converging and accurate. As briefly reviewed in Section \ref{sec:large}, our approach consists in removing large field configurations in a way that preserves the closeness to the correct answer. 

In the case of quenched $QCD$, there are several questions that are relevant for this approach and that have been addressed. How sensitive is the average plaquette $P$ to a large field cutoff \cite{effects04}? How does $P$ behave when the coupling becomes negative \cite{gluodyn04}? How does $P$ differ from its weak coupling expansion \cite{burgio97,rakow2002}? Are all the derivatives of $P$ with respect to $\beta$ continuous in the crossover region?
The analysis \cite{rakow2002,third} of the weak series for $P$ up to order 10 \cite{direnzo2000} suggests an (unexpected) singularity in the second derivative of $P$, 
or in other words in the third derivative of the free energy. In the following, we report our recent attempts to find this singularity. As all the technical details regarding 
this question have just appeared in a preprint \cite{third}, we will only summarize the 
main results leaving room for more discussion regarding the difference between series and the numerical values of $P$. 

\section{Large field configurations and perturbation theory}
\label{sec:large}

The reason why perturbation theory sometimes fail is well understood for scalar field 
theory. Large field configurations have little effect on commonly used observables but 
are important for the average of large powers of the field and dominate the large order 
behavior of perturbative series. 
A simple way to remove the large field configurations consists in 
restricting the range of integration for the scalar fields.
\begin{equation}
\prod_x \int_{-\phi_{max}}^{\phi_{max}}d\phi_x \ .
\nonumber\end{equation}
For a generic observable $Obs.$ in a $\lambda \phi^4$ theory, we have then
\begin{equation}
Obs.(\lambda )\simeq\sum_{k=0}^{K}a_k(\phi_{max})\lambda^k
 \nonumber\end{equation}
The method  produces series which apparently converge  
in nontrivial cases such as the anharmonic oscillator and $D=3$ Dyson hierarchical model 
\cite{convpert,tractable}. 

The modified theory with a field cutoff differs from the original theory. Fortunately, 
it seems possible, for a fixed order in perturbation theory, to adjust the field cutoff to an optimal value $\phi_{max}(\lambda,K)$ in order to minimize or eliminate the discrepancy with the (usually unknown) correct value of the observable in the original theory.  In a simple example\cite{optim}, 
the strong coupling can be used to calculate approximately this optimal $\phi_{max}(\lambda,K)$.  This method provides an approximate treatment of 
the weak to strong coupling crossover and we hope it can be extended to gauge theory where this crossover \cite{kogut80} is a difficult problem. 
The calculation of the modified coefficients remains a challenge, however approximately universal features of the transition between the small and large field cutoff limits 
for the modified coefficients of the anharmonic oscillator \cite{asymp}, suggest the existence of simple analytical 
formulas to describe the field cutoff dependence of large orders coefficients. 

This method needs to be extended to the case of lattice gauge theories.
Important differences with the scalar case need to be understood. For compact groups such as $SU(N)$, 
the gauge fields are not arbitrarily large. Consequently, it is possible to define a 
sensible theory at negative $\beta=2N/g^2$. However, the average plaquette 
tends to two different values in the two limits $g^2\rightarrow \pm 0$ \cite{gluodyn04}.
This precludes the existence of a regular perturbative series about $g^2=0$. 
A first order phase transition near $\beta =-22$, was also observed \cite{gluodyn04} for $SU(3)$.

The impossibility of having a convergent perturbative series about $g^2=0$ 
is well understood \cite{plaquette} in the case of the partition function 
for a single 
plaquette which after gauge fixing to the identity on three links reads.
\begin{equation}
Z=\int dU {\rm e} ^{-\beta(1-\frac{1}{N}Re TrU)}\ ,
\end{equation}
If we expand the group element $U=e^{igA}$ with $A=A^aT^a$  and the Haar measure in powers of $g$, we obtain a converging sum that allows us to calculate $Z$ accurately,  however, the ``coefficients'' are $g$-dependent.
This comes from the finite bounds of integration of the gauge fields that 
are proportional to $1/g$. If $g^2$ is small and positive, we can extend the range of integration to infinity with errors that seem controlled by $\rm{e}^{-2\beta}$. 
By ``decompactifying'' the gauge fields, we have transformed a converging sum into a power series in $g$ with constant 
coefficients growing factorially with the order. 
The situation is now resemblant to the scalar case and can be treated using this analogy.
We can introduce a gauge invariant field cutoff that is treated as a $g$ independent quantity.
For a given order in $g$, one can use the strong coupling expansion to determine 
the optimal value of this cutoff. This provides a significant improvement in regions where 
neither weak or strong coupling is adequate \cite{plaquette}. 

This program can in principle be extended to LGT on $D$-dimensional lattices, however the calculation of the modified coefficients is difficult. An appropriately modified version of the 
stochastic method seems to be the most promising for this task.
As the technology for completing this task is being developed, we will discuss several 
questions about the average plaquette and its perturbative expansion.

\section{The average plaquette and its perturbative expansion in quenched $QCD$}

\begin{figure}
\label{fig:pade}
\includegraphics[width=2.8in,angle=0]{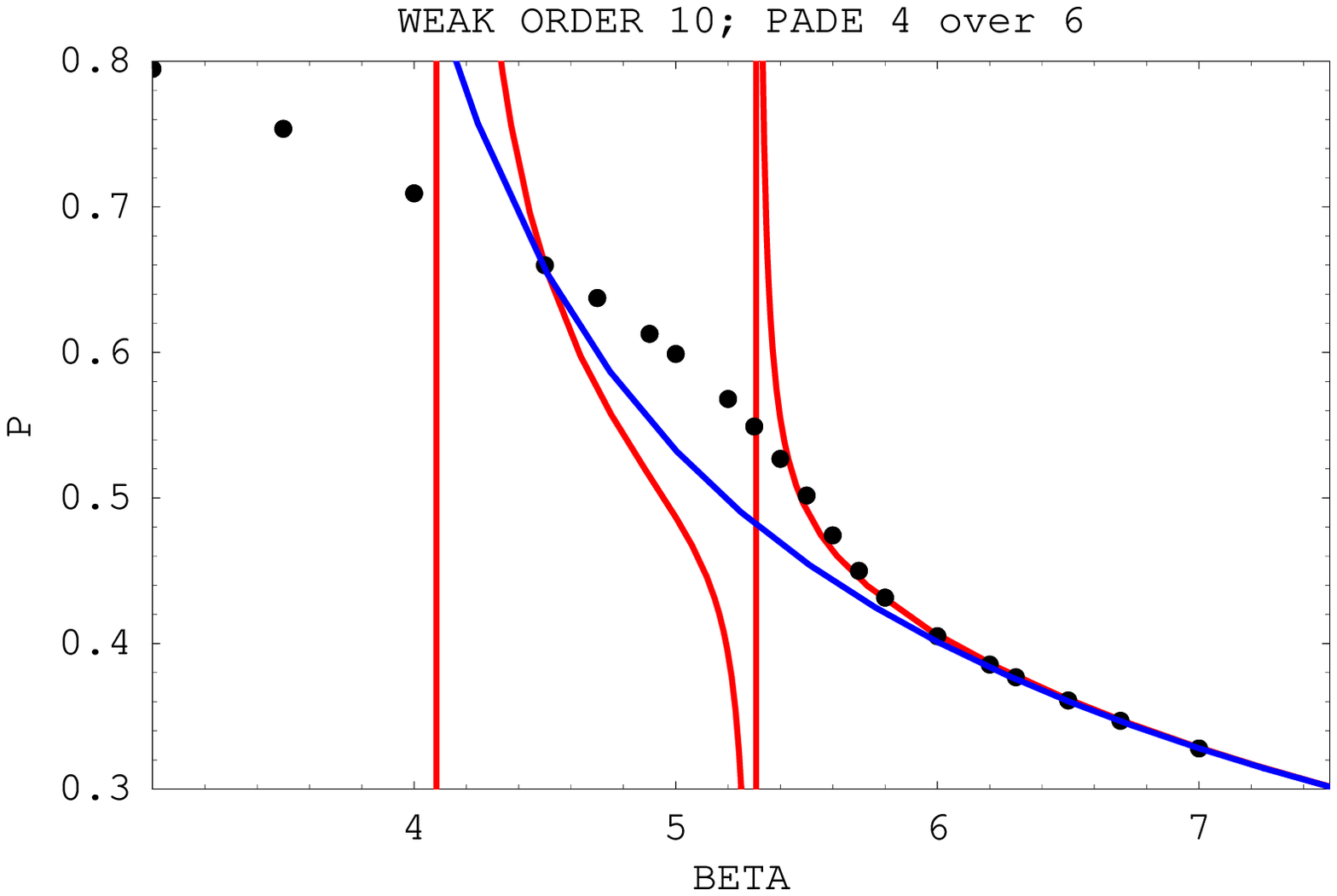}
\includegraphics[width=2.8in,angle=0]{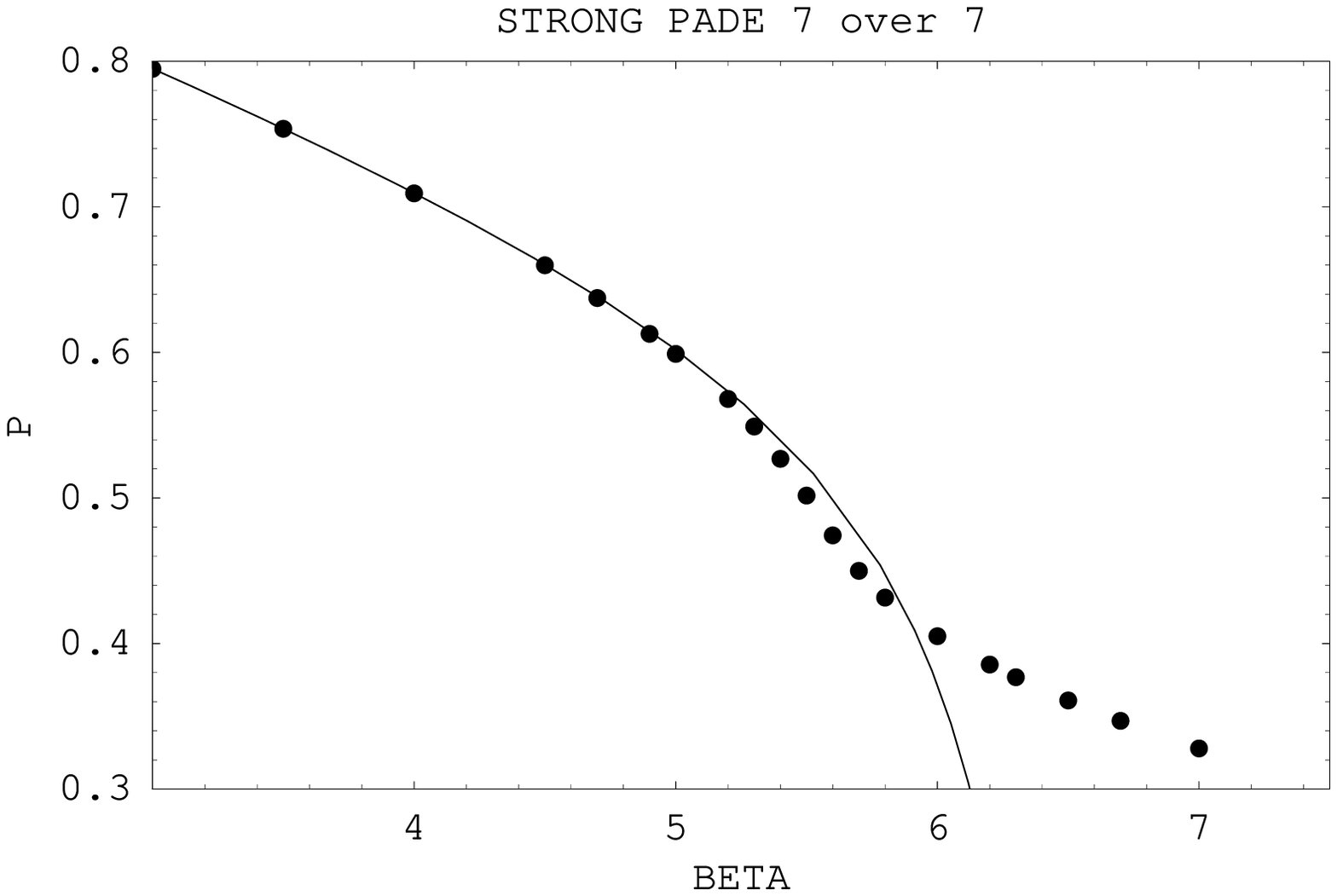}
\caption{Regular weak series (blue) and 4/6 weak Pad\'e (red) for the plaquette (left);
7/7 strong Pad\'e (right) }
\end{figure}

We now consider a $SU(3)$ lattice gauge theory in 4 dimensions without quarks (quenched $QCD$). We use the Wilson action without improvement.
Our main object will be the average plaquette action denoted $P$ and can be 
expressed as $-\partial (\rm{ln}(Z)/6L^4)/\partial  \beta$. 
The effect of a gauge invariant field cutoff is very small but of a different size below, near or above $\beta=5.6$ (see Fig. 6 of Ref. \cite{effects04}).
This is in agreement with the idea that modifying the weight of the large 
field configurations affects the crossover behavior \cite{mack78}. The weak coupling series for $P$ has been calculated up to order 10 in Ref. \cite{direnzo2000}:
\begin{equation}\nonumber
P_W(1/\beta)=\sum_{m=1}^{10} b_m \beta^{-m} +\dots.
\nonumber\end{equation}
The coefficients are given in table 1.
The values corresponding to the series and the numerical data calculated on a $16^4$ lattice 
is shown in Fig. \ref{fig:pade}. A discrepancy becomes visible below $\beta = 6$. 
The situation can be improved by using Pad\'e approximants, however, they do not show any change in curvature and often have poles near $\beta=5.2$. For comparison, Pad\'e 
approximant for the strong coupling expansion \cite{balian74err} depart visibly from the 
numerical values when $\beta$ becomes slightly larger than 5. In conclusion, it is not clear that by combining the two series we can get a complete information regarding the 
crossover behavior. 

\begin{figure}
\label{fig:apower}
\includegraphics[width=2.9in,angle=0]{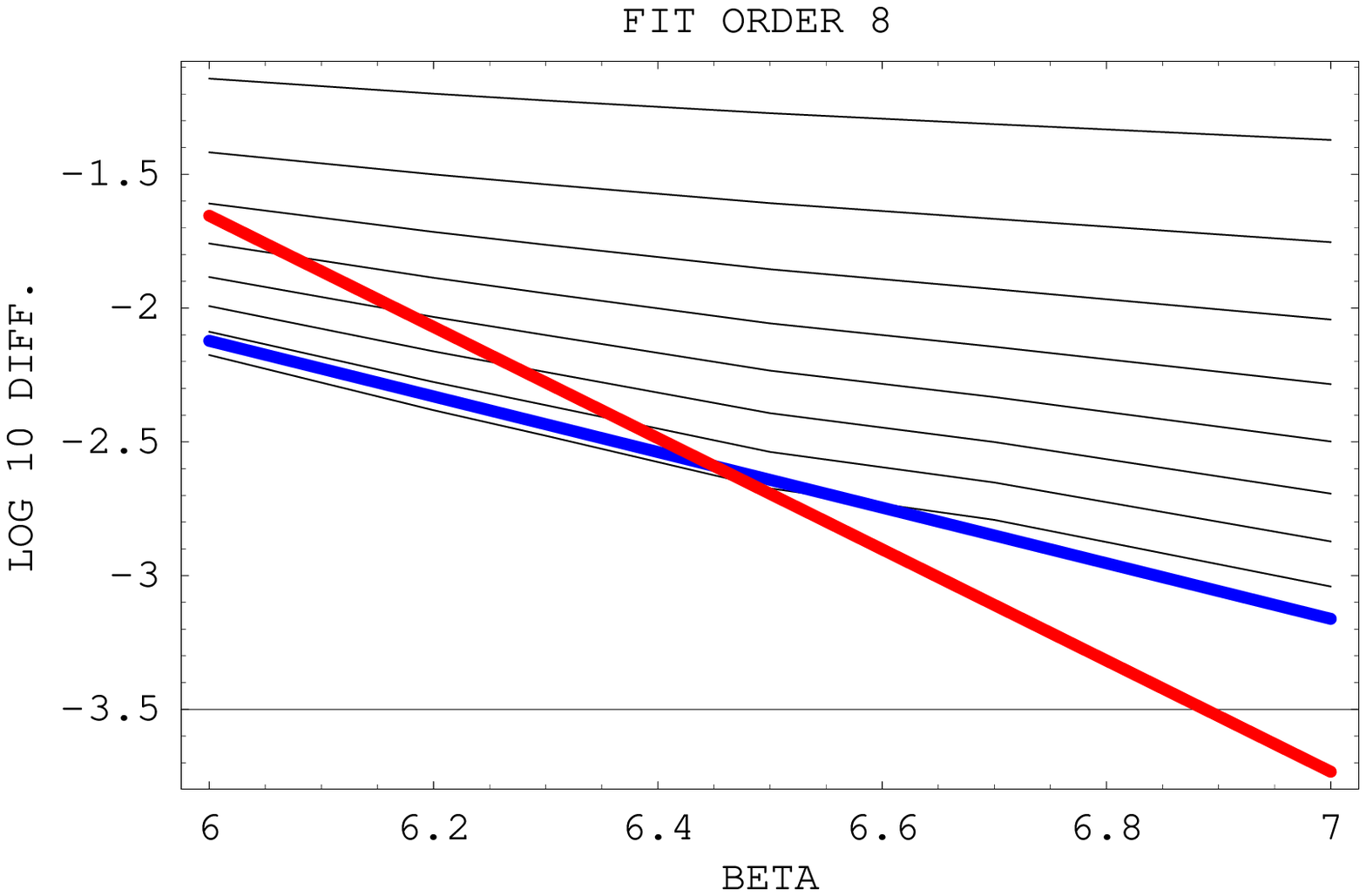}
\includegraphics[width=2.9in,angle=0]{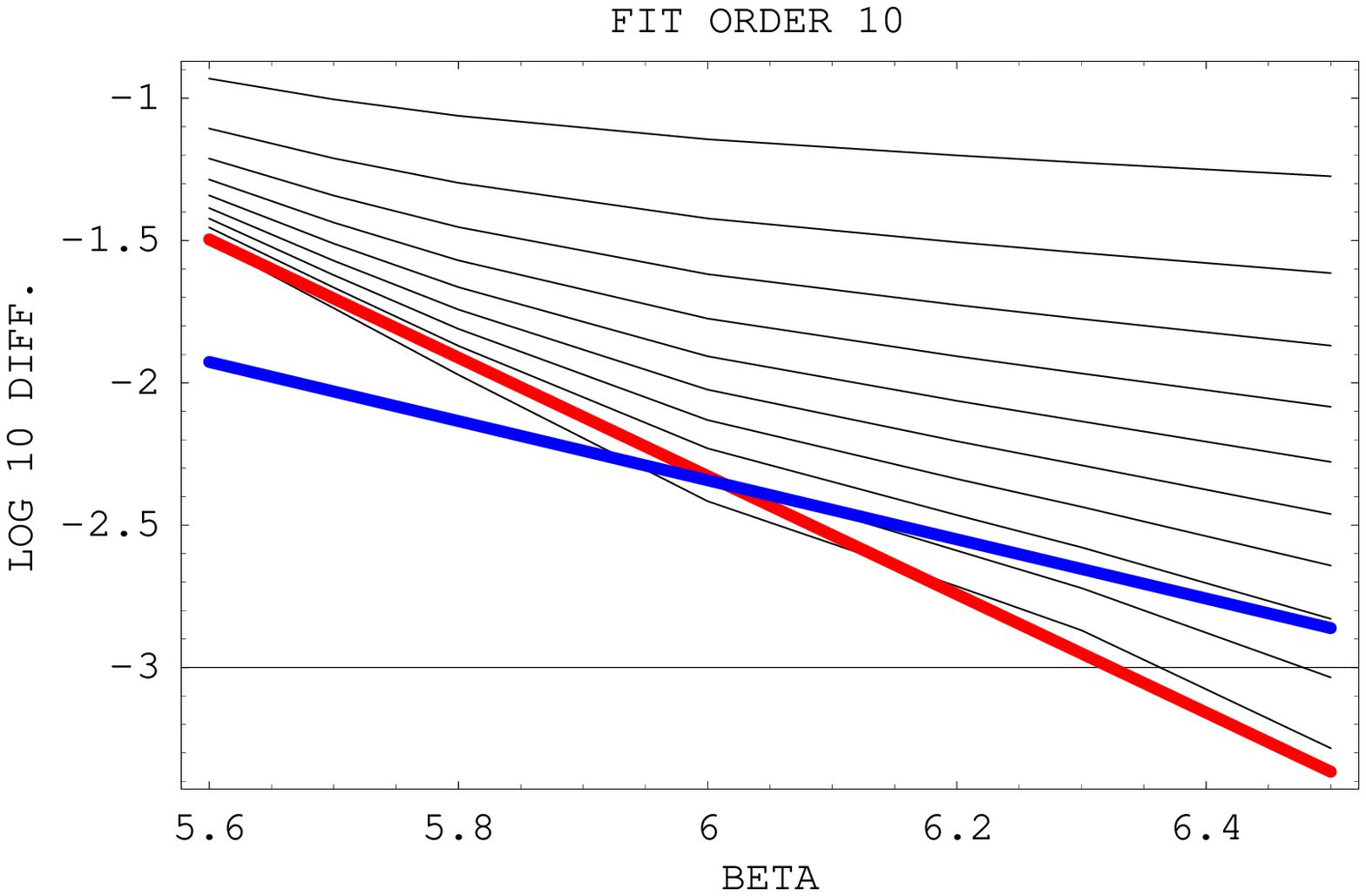}
\caption{$Log_{10}|P-P_W|$ for order 8 (left) and 10 (right, in a different range of $\beta$); the constant is fitted asumming 
$a^2$ (blue) or $a^4$ (red). }
\end{figure}

The difference between the weak coupling expansion $P_W$ and the numerical data $P$ can be further analyzed. From the example of the one-plaquette model \cite{plaquette}, one could infer that by adding the tails of integration, we should make errors of order 
$\rm{e}^{-C\beta}$, for some constant $C$. Consistently with this argument, the difference 
should scale as a power of the lattice spacing, namely
\begin{equation}
P_{Non Pert.}=(P-P_W)\propto a^A \propto \left({\rm e}^{-\frac{4\pi^2}{33}\beta} \right)^A \ .
\end{equation} 
A case for 
$A=2$ has been made in Ref. \cite{burgio97} based on a series of order 8. 
Another analysis supports $A=4$ (the canonical dimension of $F_{\mu \nu}F^{\mu \nu }$) \cite{rakow2002,rakowthese}. Fig. \ref{fig:apower} shows fits at different orders 
and in different regions that support each of these possibilities. It would be interesting to study cases where long series are available and non-perturbative effects 
well understood in order to define a prescription to extract the power properly.

The series $P_W$ has another intriguing feature:  
$r_m=b_m/b_{m-1}$, the ratio of two
successive coefficients seem to extrapolates near 6 when $m\rightarrow\infty$ when $m$ becomes large \cite{rakow2002}. This suggests a behavior of the form
 \begin{equation}\nonumber
P=(1/\beta _c -1/\beta )^{-\gamma } (A_0 + A_1 (\beta _c -\beta)^{
\Delta } +....)\ ,
\label{eq:convpar}   \nonumber\end{equation}
as encountered in the study of the critical behavior of spin models. 
We have reanalyzed \cite{third}  the series using estimators \cite{nickel80} known as the 
the extrapolated ratio ($\widehat{R}_m$) 
and the  
extrapolated slope ($\widehat{S}_m$)
in order to estimate $\beta_c$ and $\gamma $. We found that the weak series suggests
\begin{equation}
\label{eq:critical}
P\propto (1/5.74-1/\beta)^{1.08} \ .
\end{equation}
These estimators are sensitive to small variations in the coefficients and show 
a remarkable stability when the volume is increased from $8^4$ to $24^4$. The numbers are in good agreement with 
the estimates of Ref. \cite{rakow2002} with other methods.
A finite radius of convergence is 
not expected and one does not expect any singularity between 
the limits where 
confinement and asymptotic freedom hold. 
It may simply be that the series is too short to draw conclusion about its asymptotic behavior. A simple example where this happens \cite{third} is 
\begin{equation}
Q(\beta)=\int_0^{\infty}dt {\rm e}^{-t}t^{\alpha}[1-t\beta_c/(\alpha \beta)]^{-\gamma}	\ ,
\end{equation}
with $\alpha$ sufficiently large. 
If $m<<\alpha$, 
$r_m\simeq \beta_c(1+(\gamma -1)/m), $ 
For $m>>\alpha$ we have $r_m \propto m$ and the coefficients grow factorially.

If we take Eq. (\ref{eq:critical}) seriously, it implies that the second derivative of $P$ diverges near $\beta =5.7$. We have searched for such a singularity \cite{third}.
We have shown that the peak in the third derivative of the free energy present 
on $4^4$ lattices disappears if the size of the lattice is increased isotropically 
up to a $10^4$ lattice. 
On the other hand, on $4\times L^3$ lattices, a jump in the third 
derivative persists when $L$ increases. Its location coincides with the onset of 
a non-zero average for the Polyakov loop and seems consequently related to the finite 
temperature transition. It should be noted that the possibility of a third-order phase transition 
has been discussed for effective theories of the Polyakov's loop \cite{pisarski}.

A few words about the tadpole improvement \cite{lepage92} for the weak series. If we consider the resummation
\begin{equation}
P_W(1/\beta)=\sum_{m=1}^{K} e_m \beta_R^{-m} + O(\beta_R^{-K-1})
\end{equation}
with $\beta_R=\beta (1-\sum_{m=1} b_m \beta^{-m})$,  
the ratios $e_{m}/e_{m-1}$ stay close to -1.5 for $m$ up to 7, but seem to start oscillating more for large $m$.

\begin{table}[h]
\begin{tabular}{||c||c|c|c|c|c|c|c|c|c|c||}
\hline

$m$&1&2&3&4&5&6&7&8&9&10\cr
\hline
$b_m$& 2 & 1.2208 & 2.9621 & 9.417 & 
    34.39 & 136.8 & 577.4 & 2545 & 
    11590 &54160 \cr 
 $e_m$&2 & -2.779 &3.637 &-3.961 &4.766 & 
    -3.881 & 6.822 & 
    -1.771 & 17.50  & 
   48.08 \cr 
   
\hline
\end{tabular}
\caption{$b_m$: regular coefficients;  $e_m$: tadpole improved coefficients}
\end{table}

 This research was supported in part by the Department of Energy
under Contract No. FG02-91ER40664. We thank G. Burgio, F. di Renzo and P. Rakow for interesting discussions.
\providecommand{\href}[2]{#2}\begingroup\raggedright\endgroup


\end{document}